\documentstyle[a4,12pt,epsf]{article}
\newcommand{\be}{\begin{equation}}
\newcommand{\ee}{\end{equation}}
\newcommand{\bea}{\begin{eqnarray}}
\newcommand{\eea}{\end{eqnarray}}
\newcommand{\nn}{\nonumber}

\newcommand{\prd}{Phys.Rev. }

\newcommand{\prl}{Phys.Rev.Lett. }

\newcommand{\mev}{{\rm MeV}}
\newcommand{\gev}{{\rm GeV}}

\def\equiinf{\lower 0.12cm\hbox{$ \widetilde{_{_{t \rightarrow
\infty}}}$ }}
\def\equizer{\lower 0.12cm\hbox{$ \widetilde{_{_{a \rightarrow 0}}}$ }}
\begin{document}
\overfullrule 0pt

\begin{center}

{\Large{\bf HEAVY FLAVOURS LEPTONIC AND \vskip 0.3cm
SEMI-LEPTONIC DECAYS  ON THE LATTICE}}
\vspace*{2cm}

{\Large{As. Abada}}\\
\vspace*{2cm}
{Laboratoire de Physique Th\'eorique et Hautes Energies} \\
{Universit\'e de Paris XI}\\

{91405 Orsay FRANCE\footnote {Laboratoire
associ\'e au Centre National de la Recherche Scientifique.}}\end{center}
\vskip 2cm
\leftline{ABSTRACT}
\leftline{ } We present some results of the {\bf{ELC}} (European Lattice
Collaboration) study of leptonic and semi-leptonic decays of heavy-light mesons.
 Results on the decay constants $f_D, f_{D_s}$ and $f_B, f_{B_s}$, the semi-leptonic form factors of $D\longrightarrow K (K^*)$  and some preliminary estimates of form factors relevant for the $B\longrightarrow \pi , \rho$ are summarized. The $1/M$ corrections to asymptotic scaling laws are discussed. 
\begin{flushright}
 LPTHE Orsay-94/57\\ 
 hep-ph/9406352
\end{flushright}

\setlength{\parindent}{0in}
\textwidth 6.5in

{
{\bf{1- Introduction}}\vskip 0.2cm
\label{sec:intro}
 {Lattice QCD, a non-perturbative method based on the QCD Lagrangien and it's parameters, allows to predict many quantities such as hadron masses, the meson decay constants (DC), the $B$-parameter relevant for $B-\bar B$ mixing and the (SL) semi-leptonic form factors (FF) which play a crucial role in the determination of the C.K.M mixing matrix. It can be used also to test the scaling laws predicted by the HQET (Heavy Quark Effective Theory).
This method is in continuous theoretical improvement (improved actions$\dots$) and since it is based on numerical simulations, there is actually an unlimited improvement in accuracy (dedicated machines$\dots$). But it has some limitation, for instance, we are limited to compute exclusive transitions with no more than one hadron in the final state. Moreover, with present computing machines, precise predictions can be made only in the quenched approximation and the systematic errors due to this latter cannot be evaluated. There are however some results obtained in the unquenched theory
\cite{HEMCGC} suggesting that the systematic effect of quenching is small when we deal with heavy quarks. 
This is the study of weak decays of heavy mesons (composed by a heavy quark and a light one) into light ones.\vskip 0.2cm
{\bf{2- Strategy to study heavy flavours on the lattice}}\vskip 0.2cm
The heavy quarks cannot move on a lattice because their masses are larger than the natural cut-off of the theory which is the inverse lattice spacing $a^{-1}$. 
To move on a lattice, a quark has to have a mass $m\ll a^{-1}$ (in practice, $m\le 0.7 a^{-1}$). 
The present lattice spacing ranges from $2.5$ $\gev$ to $4$ $\gev$, and ELC's one is $a\simeq 3.6$ $\gev$. In the $\overline {MS}$ scheme, $m_c\simeq 1.3$ $\gev$ and $m_b\simeq 4.5$ $\gev$, so the study of the $D$ meson is easy while it is presently impossible for a physical $b$ quark to live on a lattice even the smallest lattice spacing is still larger than it's Compton wavelenght. So we are unable to study directly the $B$ meson on the lattice. Nevertheless, indirect informations may be available following this strategy:
\begin {itemize}\item{We exploit the freedom in the choice of the masses (the Wilson hopping parameter $\kappa$-Wilson), then we create light mesons, the $D$ meson and we increase the masses up to the limit $0.7 a^{-1}$, we thus have fictitious ``$D$" mesons, heavier than the $D$ but still lighter than the $B$. We call this mass region the moving quark region.}\item{On the other hand, a method proposed by Eichten \cite{eichten} allows one to put infinite mass on the lattice and the latter is considered in this approach as a static source of color.}\end{itemize}
Then one computes a physical quantity in the two mass regions (moving and static) and extracts it for the $B$ meson by interpolation with the help of the scaling laws of the HQET. The value in the static limit reduces the incertainties due to the extrapolation from the charm region. This method has shown to be very effective in the estimation of the pseudo-scalar DC. In the case of SL FF, at our large value of $\beta$ ($6.4$), our small statistics doesn't allow to compute the FF in the static limit. Nevertheless, we can study their scaling behaviour and try an extrapolation to the $b$-quark. Our prediction concerning the $B$ SL decays remain at a semi-quantitative level but it shows that the study is feasible.\\
{\bf{Lattice Setup}}: this work has been done on a $24^3 \times 60$ lattice using the standard Wilson action at $\beta=6.4$ for the gauge fields
and the quark propagators\cite{wi}, in the quenched approximation. Further details on the lattice calibration, fitting procedures, mass spectrum, extraction of matrix
elements of local operators between the vacuum and meson states, e.g.
$<M_P\vert \bar Q \gamma_5 q\vert 0>$, can be found in ref.\cite {orsw}.\vfill \pagebreak
{\bf{3- Leptonic decays $D(B)\rightarrow \ell\nu$}}\vskip 0.2cm
The mass of any hadron can be obtained from the study of an appropriate euclidean correlation function as the coefficient of its exponential time dependence:
\def\={\ =\ }
\def\x{ {\bf x}}
\def\y{ {\bf y}}
\def\0{ {\bf 0}}
\begin{equation}
{
G(t)\= \int {\rm d}^3 {\bf x}
\langle \bar u(\x ,t)\gamma_0 \gamma_5 c(\x ,t) 
\ \bar c(\0 ,0) \gamma_0 \gamma_5 u(\0 ,0) \rangle \ { _{\simeq\atop{t \rightarrow \infty}}}{f_D^2 m_D\over 2 } {\rm e}^{-m_D t} .}\label{GdeT}\end{equation}
The determination of the expectation value in eq. 
\ref{GdeT} is a non perturbative problem which can be solved numerically.
A second approach is based on the expansion of the heavy
quark (H) propagator in inverse powers of the quark mass as proposed by
Eichten \cite{eichten}; the H is static and does not live effectively
on the lattice but the quantity $f_{_H}\sqrt{M_{_H}}$ can be measured and is predicted to be independent of the heavy mass. The confrontation between the two methods is presented in Fig.1. When $M_H\rightarrow \infty$, the vector (V) and pseudoscalar (P) DC scale with the mass of the heavy quark, $M_H$, \cite{eichten}-\cite{voloshin}($M=M_P=M_V=M_H$, $\beta_0=11-{2\over 3}N_f$): 
\be {M\over f_V}=f_P={C\over \sqrt{M}}\alpha_s(M)^{-2/\beta_0}\label{echelle}\ee

$$
\epsfbox{moriondfig.ps}
$$
\parbox [t]{\textwidth} {\small {Fig.1.{\it{The $f_P\sqrt{M_P}$ is reported as a function of $1/{M_P}$. Results from other lattices are reported.
 }}}}
\vskip .2cm
A horizontal line means eq. \ref{echelle}, a slope means corrections. The curves refer to a linear ($1/M$ correction) and a quadratic fit ($1/{M^2}$ correction). The vertical line identifies the physical $B$ meson.
The Figure.1 is very illustrative, 
the first point to notice is the consistency between the moving quark results and those from the static one. It appears also that there are large corrections to the assymptotic scaling behaviour 
(\ref{echelle}). The main ELC results compared to the other
 lattice calculations and experiments are reported on Table 1 (2) concerning the $D$ ($B$) meson.
\vskip 0.2cm
\begin{table}
\centering
\begin{tabular}{|c|c|c|c|}
\hline
Ref.& $f_D$($\mev$) & Ref. &$f_{D_s}$($\mev$)\\ \hline 
ELC \cite {orsw} &$210 \pm 15$& ELC\cite {orsw} &$227 \pm 15$  \\ \hline
APE \small{(Clover)}\cite{ape} &$218\pm 9$&\small{(Clover)}\cite{ape}&$240 \pm 9$ \\ \hline
Bernard et al.\cite{bernard}&$208(9)\pm 35\pm 12$&Bernard et al.\cite{bernard}&$230(7) \pm 30 \pm 18$  \\ \hline
UKQCD\cite{ukqcd}
&$185^{+4}_{-3}$ $^{+42}_{-7}$
&UKQCD\cite{ukqcd}
&$212^{+4}_{-4} $$^{+46}_{-7}$  \\ \hline
-&-&EXP (WA75)\cite{wa75} &$232 \pm 45 \pm 20 \pm 48$  \\ \hline

-&-&EXP (CLEO2)\cite{cleo2} &$344 \pm 37\pm 52 \pm 42$ \\ \hline
 -&-&EXP (ARGUS)\cite{argus} &$267 \pm 28$\\ \hline 
\end{tabular}{\small{
\caption{\it{{Notice the agreement inside the lattice community. Taking the errors into acount, there is a fair agreement with the experimental datas (``Clover" means continum limit improved action).}}}
\label{tab:final1}}}
\end{table}
\begin{table}
\centering
\begin{tabular} {|c|c|c|c|}
\hline
Ref.& $f_B$($\mev$) & Ref. &${f_{B_s}\over f_{B_d}}$\\ \hline 
 ELC \cite {orsw} &$205 \pm 40$& ELC\cite {orsw} &$1.08 \pm 0.06$  \\ \hline

 APE {(Static-Clover)}\cite{apestat} &$290\pm 15 \pm 45$&  \small{(Static-Clover)}\cite{apestat}&$1.11(3)$ \\ \hline 

 UKQCD \small{(Static)}\cite{ukqcd} &$253 ^{+16}_{-15}$ $^{+105}_{-14}  $& \small{(Static)}\cite{ukqcd}&$1.14^{+4}_{-3}$ \\ \hline

 Bernard et al.\small{(Static)}\cite{bernard}&$235(20)\pm 21\pm 12$& Bernard et al.\small{(Static)}\cite{bernard}&$1.11 \pm 0.05 $  \\ \hline
HEMCGC \small{(Unquenched)}\cite{HEMCGC} &$200\pm 48$ && \\  \hline
\end{tabular}{\small{
\caption{\it{{Notice that the HEMCGC unquenched result is not different from those using the quenched approximation (``Static" means infinite mass limit). }}}
\label{tab:final2}}}
\end{table}
\vskip 0.2cm
{\bf{ {The $B-$ parameter}}}: the physical predictions for the $B-\bar B$ mixing depend on the value of the $B-$parameter of the heavy light $\Delta B =2$ four quark operator \cite{orsw}:
\be 
{B_{_{D^0}}} = 1.05\ \pm \ 0.08,\quad
B_{_{B^0}} = 1.16\ \pm \ 0.07,\quad
{B_{_{D_s}}\over B_{_{D_d}}} \simeq {B_{_{B_s}}\over B_{_{B_d}}}=
1.02 \pm  0.02 \ee
Concerning the experimental implications, with the results quoted above,
we can predict: 
${
f_{B_d}\ \sqrt{{\rm B}_{B_d}}\= 220\ \pm\ 40\ \mev
}$ which is the combination relevant for the $CP$ violating effects in $B$ decays and for $B-\bar B$ mixing amplitude.\vskip 0.2cm
{\bf{4- Semi-leptonic decays $D(B)\rightarrow K,K^* (\pi, \rho) \ell\nu$}}\vskip 0.2cm
From the study of three-point correlation functions
\cite{victor}, one extracts the weak current matrix elements
for a given momentum transfer $q$:
\be < K \vert J_\mu \vert D > = \Bigl( p_D + p_K -
\frac{ M_D^2 - M_K^2}{q^2} q \Bigr)_\mu f^+_K(q^2) +
\frac{ M_D^2-M_K^2}{q^2} q_\mu f^0_K(q^2) \label{dk}\ee
\bea
 < K^*_r \vert J_\mu \vert D > &=& e^\beta_r \Bigl[ \frac{2
V(q^2)}{M_D+M_{K^*}}
\epsilon_{\mu\gamma\delta\beta}p_D^\gamma p_{K^*}^\delta + i 
( M_D +M_{K^*}) A_1(q^2) g_{\mu\beta} \nn \\ &-& i
\frac{A_2(q^2)}{M_D+M_{K^*}}P_\mu q_\beta + i \frac{A(q^2)}{q^2}2
M_{K^*}q_\mu P_\beta \Bigr] \label{dks} , \eea $q=p_D-p_{K(K^*)}$, $P=p_D+p_{K(K^*)}$ and $e^\beta_r$ is
the polarization vector of the $K^*$. $f^{+,0}_K$, $V$, $A_{1,2}$, $A$
are dimensionless FF \cite{lubicz}. By varying the Lorentz component of the current (\ref{dk}), (\ref{dks}), the meson momenta and the $K^*$
polarization, we can extract the 6 FF (only 4 are relevant in the decay rates). As said before, our study tries to extrapolate the FF to $B\rightarrow \pi,\rho$ decays using the scaling laws predicted by the HQET \cite{scala},\cite{pene}. On another hand, we have studied the $q^2$ dependance by putting a set of discret (volume is finite) momenta on the lattice for the final meson (the initial being at rest): $\vec {p_k}= $ $(0,0,0)$ (for which $q^2=q^2_{max}$), $(1,0,0)$, $(2,0,0)$, $(1,1,0)$ and $(1,1,1)$ (in lattice unit). We have computed
the matrix elements (\ref{dk}, \ref{dks}) and extracted the FF for different
 $q^2$. All the computational details can be found in ref.\cite{semi}. At $\vec {p_k}=(1,0,0)$, we are near $q^2=0$ where we have the maximum of the partial widths of the $D$ meson, but for the $B$, we are very far from $q^2=0$, then we had to perform an extrapolation to $q^2=0$, and the best that we could do was to use the nearest pole dominance\cite{pene}: $FF(q^2)={FF(0)\over 1-q^2/M_t^2}$, $M_t$ being the mass of the meson exchanged in the t-channel.\vskip 0.2cm
{\bf{ \underline{$D$ meson FF}}}: the main ELC results compared to the other
 lattice calculations, theoretical models and experiments are reported on Table 3. 

\begin{table}
\centering
\begin{tabular}{|c|c|c|c|c|}
\hline
Ref& $f^+_K(0)$ & $V(0)$ & $A_1(0)$&$ A_2(0)$\\ \hline
 ELC.\cite{semi} &$0.65
\pm 0.18$&$0.95 \pm 0.34$&$0.63 \pm 0.14$& $0.45 \pm 0.33$ \\ \hline

Lat.\cite{victor}&$0.63 \pm 0.08$&$0.86 \pm 0.10$&$0.53 \pm
0.03$& $0.19 \pm 0.21$ \\ \hline

Lat.\cite{bes}&$.90\pm.08\pm .21$&$1.43\pm .45\pm .49$&$.83\pm
.14\pm .28$& 
$.59 \pm .14\pm .24$\\ \hline
QM.\cite{wsb}& $0.76$&$1.23$&$0.88$&$1.15$\\ \hline
QM.\cite{wisg}& $0.76-0.82$&$1.1$&$0.8$&$0.8$ \\ \hline
SR.\cite{aos}& $0.6\pm 0.10$&$-$&$-$&$-$ \\
\hline 
SR.\cite{bbd}& $0.6^{+0.15}_{-0.10}$&$1.1\pm 0.25$&$0.5\pm0.15$&$0.6\pm0.15$ \\ \hline 
Exp.\cite{e691}&$.70 \pm .08$&$.9 \pm .3\pm .1 $&$ .46 \pm .05\pm 05
$&$0.0 \pm 0.2\pm 0.1 $\\
\hline 
\end{tabular}{\small{
\caption{{\it{$D \rightarrow
K$ and $K^*$ FF. ``Lat.'' refers to lattice QCD, ``QM'' to quark models, ``SR''
to QCD sum rules and ``Exp.'' to experiment.}}} \label{tab:final}}}
\end{table}

\vskip 0.2cm
{\bf{\underline{$B$ meson FF}}}: the HQET suggests that, when $M\rightarrow \infty$ at fixed $\vec
p_{K,K^*}$ and with $\vec
p_{K,K^*}\ll M$, the FF scale as \cite{scala},\cite {pene}:
$f^+= M^{1/2} \, \gamma_{+}\times\Bigl( 1 + \frac{\delta_{+}}{ M} \Bigr)$, 
$V = M^{1/2} \, \gamma_{V} \times\Bigl( 1 + \frac{\delta_{V}}{ M} \Bigr)$, 
$A_2 = M^{1/2} \, \gamma_2 \times\Bigl( 1 + \frac{\delta_{2}}{ M} \Bigr)$ and 
$A_1 = M^{-1/2} \, \gamma_1 \times\Bigl( 1 + \frac{\delta_{1}}{ M}\Bigr)$. The coefficients $\delta$ and $\gamma$ are summarized in Table 4.
\begin{table}
\centering
\begin{tabular}{|c|c|c|c|c|}
\hline
$\vec p$ &$\gamma_+ \, \gev ^{-1/2}$ & $\gamma_V \, \gev ^{-1/2}$ &
$\gamma_1 \,\gev ^{+1/2}$& $\gamma_2 \, \gev ^{-1/2}$\\ \hline
$(0,0,0)$ &$-$&$-$&$0.96 \pm 0.16$& $-$ \\ \hline $(1,0,0)$ &$0.39 \pm
0.25$&$0.29 \pm 0.12$&$1.05 \pm 0.25$& $0.44 \pm 0.25$ \\ \hline \hline $\vec
p$& $\delta_+\gev $  & $\delta_V\gev$ & $\delta_1\gev$ &$\delta_2\gev$ \\ \hline
$(0,0,0)$ &$-$&$-$&$-0.33 \pm 0.09$& $-$ \\ \hline $(1,0,0)$ &$0.0 \pm
1.1$&$1.9 \pm 1.3$&$-0.46 \pm 0.22$& $-0.6 \pm 0.8$ \\ \hline
\end{tabular}
\caption{\it{The coefficients of the $1/M$ expansion of the FF.}}
\label{tab:scale}
\end{table}

In Figure.2, we have plotted the FF as a function of ${1\over M}$.
\vskip 0.2cm
{\bf{5- Conclusion}}\vskip 0.2cm
After a first study of the leptonic heavy meson decays, we have looked at the semi-leptonic ones;
we have done, for the first time on the lattice, the study of the scaling law predicted by the HQET concerning the FF: we get satisfactory quantitative results for the $D$ meson. Concerning the $B$ meson, we have good results near $q^2_{max}$ and the results we get at $q^2=0$ have very large errors, moreover, our predictions depend on the pole dominance approximation (more theoretical knowledge on $q^2$ behaviour of the FF is needed). But these results should be taken as a first indication of the feasability of this method. Finally, $1/M$ corrections to the scaling laws (leptonic and semi-leptonic decays) predicted by the HQET are large.
\setlength{\unitlength}{1mm}
$$
\epsfbox[120 150 400 460]{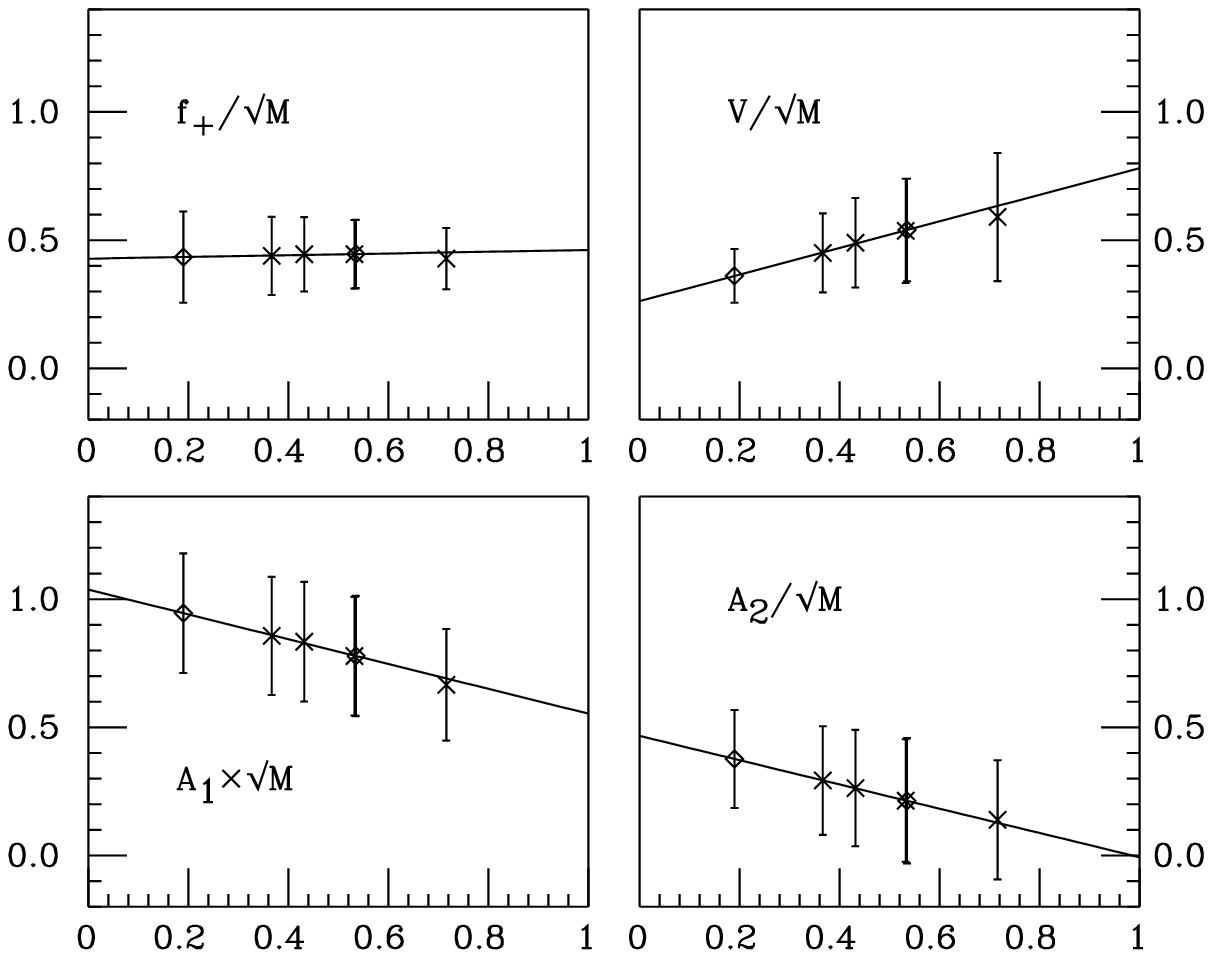}
$$
\parbox [t] {\textwidth}{{Fig.2.{\it{\small{The crosses are the lattice points and the diamonds are the extrapolation to the $D-$ and $B-$meson.} }}}}\vskip 1cm

\noindent {\bf Acknowledgments :} \par
We thank the organisers of the ``Rencontres de Moriond", particularly Madame Jo\"elle Raguideau for her efficient and friendly work.
 This work was supported in part by the CEC Science Project 
SC1-CT91-0729.\par

\end{document}